\documentclass{sig-alternate}

\usepackage[english]{babel}
\usepackage[utf8]{inputenc}
\usepackage{amsmath}
\usepackage[table,xcdraw]{xcolor}
\usepackage{graphicx}
\usepackage{enumitem}
\usepackage[caption=false]{subfig}
\usepackage{mwe}
\usepackage[colorinlistoftodos]{todonotes}
\usepackage{url}
\usepackage{tabularx}

\usepackage{multirow}

\usepackage{flushend}	% balance columns on last page
\usepackage[bookmarks, pdftex, colorlinks=false]{hyperref}     % clickable references
\usepackage[square,numbers,sort]{natbib}     % better references

\graphicspath{{figures/}}

\begin{document}
	%
	% --- Author Metadata here ---
	%\conferenceinfo{ACM Multimedia}{2013 Barcelona, Spain}
	%\CopyrightYear{2007} % Allows default copyright year (20XX) to be over-ridden - IF NEED BE.
	%\crdata{0-12345-67-8/90/01}  % Allows default copyright data (0-89791-88-6/97/05) to be over-ridden - IF NEED BE.
	% --- End of Author Metadata ---
	
	\title{Where to be wary: The impact of widespread photo-taking and image enhancement practices on users' geo-privacy}
	%Protective Post-processing: Uploader enhancements for cloaking the geo-location of social images

	\author{
			Jaeyoung Choi$^{1,2}$, Martha Larson$^2$, Xinchao Li$^2$, Gerald Friedland$^1$, Alan Hanjalic$^2$ \\\\
			\affaddr{$^1$International Computer Science Institute, Berkeley, CA, USA} \\ 
			\affaddr{$^2$Delft University of Technology, Delft, Netherlands} \\\\
			\email{$^1$\{jaeyoung, fractor\}@icsi.berkeley.edu}\\
			\email{$^2$\{m.a.larson, x.li-3, a.hanjalic\}@tudelft.nl}\\
		}
	
	\maketitle
	\begin{abstract} 
	%Today, turning off a camera's GPS does not guarantee that the location at which a photo was taken cannot automatically be predicted. 
		Today's geo-location estimation approaches are able to infer the location of a target image using its visual content alone.
		These approaches exploit visual matching techniques, applied to a large collection of background images with known geo-locations. 
		Users who are unaware that visual retrieval approaches can compromise their geo-privacy, unwittingly open themselves to risks of crime or other unintended consequences.
		Private photo sharing is not able to protect users effectively, since its inconvenience is a barrier to consistent use, and photos can still fall into the wrong hands if they are re-shared.
		This paper lays the groundwork for a new approach to geo-privacy of social images: Instead of requiring a complete change of user behavior, we investigate the protection potential latent in users existing practices.
		%, or of the consequences of publicly uploading images that are not geo-protected. 
		%Unawareness contributes to the larger issue of users being unmotivated to protect themselves. 
		We carry out a series of retrieval experiments using a large collection of social images (8.5M) to systematically analyze where users should be wary, and how both photo taking and editing practices impact the performance of geo-location estimation.
		We find that practices that are currently widespread are already sufficient to protect single-handedly the geo-location (`geo-cloak') up to more than 50\% of images whose location would otherwise be automatically predictable.
		Our conclusion is that protecting users against the unwanted effects of visual retrieval is a viable research field, and should take as its starting point existing user practices.

	\end{abstract}
	\section{Introduction}
	% \begin{itemize}
	% \item Explain what the paper is about
	% \item Explain automatic geo-location estimation (introduce the terminology: background collection "aka database")
	% \item Explain why it is important: include Collections like Google street view are making this more likely
	% \item Motivate the methodology
	% \end{itemize}
	\begin{figure}[!t]	
		\includegraphics[clip,width=\columnwidth]{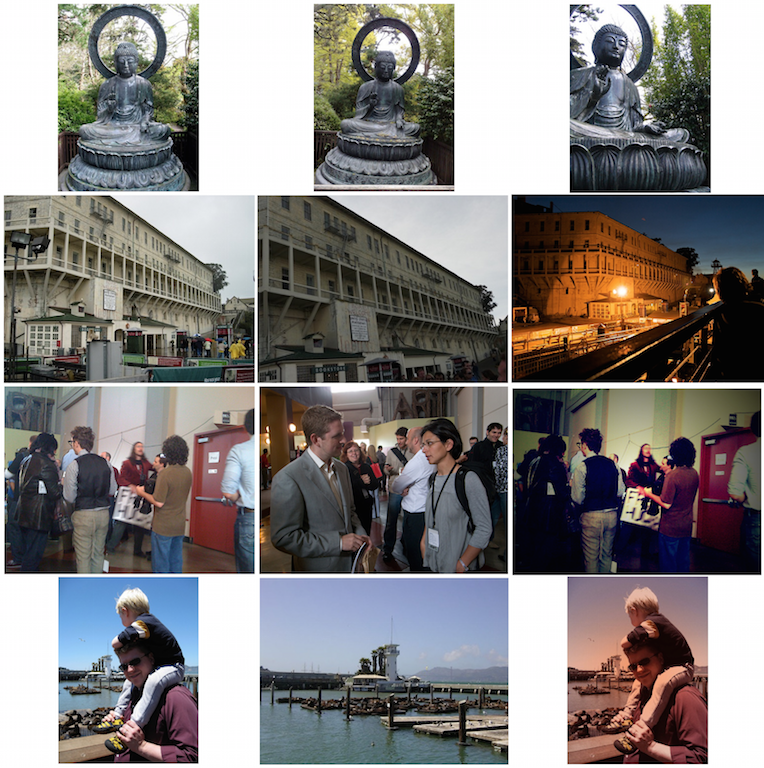}	
		
		\caption{Geo-location Estimation can automatically infer the location of a target photo, shared online by a user (first column), if at least one visually matching `geo-propagator' image exists in the background collection (middle column). However, widespread  practices, such as framing (last column, top two rows) or enhancement (rightmost column, bottom two rows) can serve to protect image location information.}
		
		\label{fig:firstimage}
		
	\end{figure}
	As image technologies improve, it is possible in an increasing number of cases to automatically infer the location at which a photo was taken on the basis of its visual content.
	Most work on Geo-location Estimation (GLE), emphasizes the usefulness of geo-coordinates in organizing photo collections, and improving multimedia retrieval e.g.,~\cite{IM2GPS, Serdyukov:2009, VanLaere2012, Trevisiol:2013}.
	However, the success of GLE in supporting users in browsing and finding photos, depends critically on our ability as a multimedia research community, to understand and mitigate its risks.
	%Specifically, we must work to prevent users from making themselves vulnerable by acts of image sharing whose implications they do not understand.
	Users unknowingly release information about their location when they share photos, surrender control of their own geo-privacy, and make themselves vulnerable.
	The purpose of this paper is to lay the foundation for a new line of research directed at supporting users in protecting themselves, given their own natural phototaking and sharing behavior.
	
	The key points about GLE for social images that are relevant for this paper are illustrated by the examples in Fig.~\ref{fig:firstimage}. 
	The first image of each row is a photo that has been uploaded by a user to a popular photosharing platform (Flickr), and whose geo-location is correctly predicted by our GLE approach.
	Our approach, which uses geometric visual matching, represents the state of the art in GLE, and is described in detail later.
	At this point, we introduce its core mechanism, which is typical for GLE technology for social images.
	The approach makes use of a large background collection containing photos for which the geo-location is already known, for example, because the photos are accompanied by GPS coordinates or manually geo-tagged by the uploading user.
	Here, we use 8.5 million images.
	The target image is used to retrieve `geo-propagators', visually matching images from the background collection.
	In Fig.~\ref{fig:firstimage}, the first image of each row is a target image, i.e., a photo that has been shared by a user.
	Geo-location prediction is accomplished by propagating the geo-coordinates of one or more geo-propagators to the target image. 
	
	In Fig.~\ref{fig:firstimage}, the middle image of each row is the image that was found by our GLE system and used as a `geo-propagator' to infer the location of the target image. This image was taken by a different user, who either used a camera that recorded GPS coordinates, or provided the geo-location during upload. Note that GLE does not build a explicit model of a location, rather, a single photo uploaded by another user is sufficient to identify the geo-location at which the target photo was taken. In the top two rows, the target photo and the geo-propagator depict the same object/landmark. However, the visual match can also involve what could otherwise be considered unimportant background detail, as in c and d. 
	
	It is natural for the first two columns of Fig.~\ref{fig:firstimage} to be a cause for some alarm. These are photographs uploaded by users, who most likely assume that the location of the photo would be protected if they turned off the GPS of their cameras. The unintended consequences range from annoying (e.g., for newly weds who were not planning to reveal their secret honeymoon destination) to downright dangerous (e.g., for a young mother targeted by a stalker). This alarm is the factor that triggered this paper. However, the basic insight that we build on is fundamentally a positive one: Users' photo taking and sharing behavior does not always lead to geo-privacy risks. If we can analyze natural behavior that protects privacy, then we take a step towards achieving our ultimate, long term goal, of developing multimedia technology that puts control over geo-privacy back into the hands of users.
	
	The third column of Fig.~\ref{fig:firstimage} illustrates widespread practices that are naturally protective of geo-privacy. These images are comparable to those in the first column from the user perspective. However, in contrast to the first-column images, their locations can \emph{not} be predicted by our GLE algorithm. In the top two rows, we see that a user has changed the framing, i.e., the way in which the photo was taken. In moving away from the `typical' shot, the user is attempting to create a unique effect, and at the same time (and we assume without explicit intention) creates a geo-privacy protected photo. In the other two images Instagram filters have been applied: Lomo (people at conference) and Toaster (father with kid). After application of these filters, the locations of the images can no longer be correctly predicted by our GLE algorithm. 
	
	In short, these examples point towards the existence of simple practices applied by users that naturally \emph{geo-cloak} their images, i.e., prevent the locations of those images from being automatically predicted. The implication is that it is not necessary to introduce new, never-before-seen technologies in order to protect users' geo-privacy. Such technologies require user acceptance and widespread uptake before they can be effective. Rather, we can target new information retrieval technologies that build on behaviors that users naturally engage in, independently of privacy considerations. This paper represents a first step on the road to such technologies.
	
	We take the position that GLE technology has both positive and negative implications for users. For this reason, we necessarily approach the topic of geo-privacy for social images, not with the goal of developing the ultimate privacy protecting algorithm, but rather, with the goal of attaining the \emph{correct balance} between positive and unintended consequences. It is important to note that we are \emph{not} asserting that every image must be geo-privacy protected/geo-cloaked. Indeed, there are many images taken and shared by users precisely for the reason of revealing geo-location, apparently to as many other people as possible. We do not propose that every picture of the Eiffel Tower or of the Taj Mahal should be visually distinctive. Rather, we strive to understand how users' behavior impacts geo-privacy because of its importance in situations where geo-location information is `incidental' to images, and represents a potential, unintended geo-privacy leak. Because of the rate at which photos are taken and shared by users online the balance that we strive to achieve is necessarily a \emph{dynamic balance}. A photo whose geo-location cannot be identified by our system today, might well be identifiable tomorrow, as more images are taken and added to the background collection. We address the dynamic balance by considering not only the geo-cloaking of individual images, but also the connection of how the ability of a given photo to compromise the geo-privacy of the user is related to the availability of other photos online.
	
	The paper makes the following contributions:
	\begin{itemize}[noitemsep]
		\item User photo taking behavior: Experiments showing that where, how, and when users take photos impacts automatic GLE,
		\item Image editing behavior: Experiments showing the way in which widespread image enhancements (e.g., filters and cropping) impact automatic GLE,
		\item Research challenges: Jump-start a new line of research devoted to visual geo-privacy that protects against unwanted affects of visual retrieval by leveraging natural user behavior.
	\end{itemize}
	
	In the next section, we provide additional motivation for the paper and also cover related work. Then we present the results of our experiment-based analysis. We distill the results into a series of recommendations for user behaviors that promote geo-privacy protection. 
	
	\section{Motivation and related work}
	In this section, we provide additional background and motivation for our work, and discuss its relationship to existing work.
	
	\subsection{The importance of geo-privacy}
	Privacy involves the ability to control information about oneself, and, by extension, we define \emph{geo-privacy} as the ability to control information about one's location, both in the past and in the future.
	As mentioned in the introduction, leaks of personal geo-information can be annoying. The literature, however, tends to cover the more dangerous consequences, especially the phenomena of `cybercasing', burglars automatically harvesting information about potential victims online~\cite{Friedland:2010}.
	The phenomenon of stalking has been connected to opportunity~\cite{stalk:12}, and, as such, geo-information in the hands of the perpetrator renders the victim more vulnerable.
	
	In the real world, there are no failsafe protections against burglars and stalkers.
	Instead, people protect themselves by trading off convenience with conventional protection measures.
	Our goal in seeking to protect geo-privacy is to increase the number of  measures at users' disposal to protect themselves.
	
	There are obvious parallels between concerns about automatic face detection algorithms compromising privacy~\cite{deepface}, and geo-location estimation algorithms compromising privacy.
	However, geo-location estimation is potentially more pernicious, since users are likely unaware of conditions under which a GLE algorithm can automatically predict the locations of their images. 
	The first `GLE condition' is the visual distinctiveness of the image content. 
	The visual elements that allow GLE to geo-locate the images in the first column of Fig.~\ref{fig:firstimage} range from fairly obvious to barely noticeable.
	The second `GLE condition' is the presence of `geo-propagators' in the background collection that are visually closer to the target image than images taken at other locations.
	Although a user may have a rough idea of the number of photographs taken at a given location, there is no way to guess the existence of visually similar locations.
	The impossibility of deciding whether the GLE conditions are satisfied, mean that users are, an can be expected to remain, easily susceptible to inadvertently leaking geo-information when sharing photos. 
	
	%amount of information revealed is made evident by academic work: \cite{Zheng:2015} (Also the work about account matching)
	% textual approaches extending to visual approaches?
	
	Two points characterize the nature of the negative impact that automatic GLE of social images poses to user geo-privacy.
	First, GLE is dangerous not because it is able to identify the location at which an image is taken, but rather because it is able to do so \emph{automatically}. Automatic processing is fast, and makes it possible to mine large collections of social images in search of vulnerabilities, as with cybercasing, mentioned above. 
	Second, automatic processing can potentially identify location in cases which would be challenging for a human observer. The number of people who can identify the locations of the images in Fig.~\ref{fig:firstimage} is relatively small. They can, however, be identified by anyone with access to GLE technology. 
	The goal in protecting geo-privacy is not to completely eradicate evidence of location in social images, but rather to slow down or hold back the ability of GLE algorithms to automatically process large quantities of social images and extract location information.
	Users should retain control over whether or not their images are geo-locatable or not.
	
	\subsection{Social images and geo-location estimation}
	
	This paper takes the position that it is important to understand how user behavior impacts automatic geo-location estimation technology, since this technology opens users to risks that they may be unaware of.
	It should be noted that this is a nuanced stance---we do not claim that all automatic geo-location estimation is bad, rather that we need to understand the extent to which users can control whether or not it can be applied to their photos.
	In this subsection, we briefly review the work on geo-location estimation. This work has generally emphasized the positive side of image GLE, and the potential it has to serve users in browsing and finding images.
	
	\subsubsection{Text-based image Geo-location Estimation}
	
	An important body of work on automatic geo-location estimation for social images exploits textual information in the form of tags and descriptions that are contributed by users, usually the uploaders themselves. 
	Early work included~\cite{Serdyukov:2009}, and ultimately the most effective approach was one that built language models of tags, such as~\cite{VanLaere2012}.
	Anticipating the results of the experimental section, visual-based GLE does not achieve the same level of performance that has been reported for the state-of-the-art text-based GLE system using the same dataset~\cite{USEMP2013}.
	% 	Such a system uses the tags and descriptions associated by the user with images in order to estimate the location at which the image is taken. 
	% 	Notice that text-based geo-location does not dependent on users tagging photos with explicit place names, such as \emph{Berlin} or \emph{Cairo}. 
	% 	Rather it leverages general associations between words and location.
	We consider text as less of a threat to users' geo-privacy, since 
	%unlike the visual content of the image, 
	it is relatively easy to disassociate text from an image. 
	Although this paper is devoted to protecting users against the unintended consequences of visual GLE, we point out that ultimately researchers need to address the unintended consequences of hybrid textual/visual GLE.
	
	\subsubsection{Visual-based Geo-location estimation}
	%Xinchao can you please make this paragraph of text better and add the key references (just 1-2 seminal references, and then the most important of the most recent work)? 
	Visual-approaches to automatic geo-location estimation can be divided into two approaches. 
	The first set attempts to predict the geo-location given a relatively restricted range of candidate locations. 
	The second set, to which the work in this paper belongs, attempts to predict the geo-location of photos any where on the surface of the world. 
	For this purpose, retrieval techniques applied to a large background collection of geo-tagged images are exploited, as mentioned in the introduction. 
	
	The first set of Geo-constrained location prediction approaches has included city-scale landmark recognition~\cite{landmarkidentification2011}, estimation of location in target images using geo-informative visual attributes learned from training images of cities~\cite{imagRetrilocalization:2010}~\cite{fang2013giant}, or as a classification~\cite{PerLocationSVM}. 
	
	%Chen et al.~\cite{landmarkidentification2011} investigated the problem of city-scale landmark recognition using 1.7 million perspective images. A vocabulary-tree-based retrieval scheme based on SIFT features was built to approach this task. Kalantidis et al. mined representative scenes from the geo-tagged photos from $22$ European cities and then proposed an approach to estimate the location depicted in the target image by matching it with these representative scenes. Fang et al.~\cite{fang2013giant} mined the geo-distinctive image patches for each city, and incorporate the learned geo-informative visual attributes into the location recognition problem to improve the classification performance. These learned geo-informative visual attributes are shown to be useful for city-based location recognition. Gronat et al.~\cite{PerLocationSVM} tried to attack this city-scale location recognition problem from the classification point of view. They modeled each geo-tagged image in the collection as a class, and learned a per-example linear SVM classifier for each of these classes with a calibration procedure that makes the classification scores comparable to each other. Due to the high computational cost in both off-line learning and online querying phases, the experiment was conducted on a limited dataset of $25k$ photos from Google Streetview taken in Pittsburgh, U.S., covering roughly an area of $1.2 \times 1.2 km^2$.
	
	Compared to the effort that has been devoted to geo-constrained location prediction, there has been relatively less work dedicated to predicting locations at the global scale. A major contribution in this direction was the approach by Hays and Efros~\cite{IM2GPS}. They deployed various global visual representations to model the visual scene similarity between images and employed the Mean Shift Clustering approach to estimate the location. Further contributions can be found among the submissions to the Placing Task of the MediaEval multimedia evaluation benchmark, which addressed the challenge of location prediction of social images~\cite{PlacingWorkNote2013}. One of the state of the art system is the Geo-Visual Ranking approach proposed by Li et al.~\cite{TMMGVR}. Given a query image, a geo-location is recommended based on the evidence collected from images that are not only geographically close to this geo-location, but also have sufficient visual similarity to the query image within the considered image collection. This makes it to improve over two major classes of previous approaches, addressing the disadvantages of both 1-NN and clustering~\cite{IM2GPS}.
	
	% [landmarkidentification2011] D. Chen et al. City-scale landmark identification on mobile devices. In Proc. CVPR '11, 2011.
	% [imagRetrilocalization:2010] K. Yannis et al. VIRaL: Visual image retrieval and localization. Multimedia Tools and Applications,51:555{592, 2011.
	% [fang2013giant] Q. Fang, J. Sang, and C. Xu. Giant: Geo-informative attributes for location recognition and exploration. In Proc. MM' 13, 2013.
	% [PerLocationSVM] P. Gronat, G. Obozinski, J. Sivic, and T. Pajdla. Learning and calibrating per-location classiers for visual place recognition. In Proc. CVPR '13, 2013.
	% [IM2GPS] J. Hays and A. Efros. IM2GPS: estimating geographic information from a single image. In Proc. CVPR '08, 2008.
	% [TMMGVR] Li, X., Larson, M., and Hanjalic, A. Global-Scale Location Prediction for Social Images Using Geo-Visual Ranking. Multimedia, IEEE Transactions on, 17(5), 674-686.
	
	\subsection{Weakness of other geo-privacy approaches}
	
	Next we cover possible alternative approaches that could be taken to geo-privacy. 
	We mention them here in order to highlight their weaknesses, and underline the importance of gaining insight into how users natural behavior can already serve to protect their privacy.
	
	\subsubsection{Keypoint Removal and Re-injection}
	
	The forensics and image analysis communities have devoted quite a bit of effort to techniques that conceal images from SIFT-based image retrieval system by removing keypoints and forging in new keypoints ~\cite{Hsu:2010, Do:2010, Costanzo:2014in}. 
	These systems were studied in the context of copyright detection and explore the robustness of near duplicate detection systems. 
	This work differs from our own in that social images uploaded by users are very diverse, and for this reason, the need to handle near-duplicates is not common in GLE. 
	
	There are two drawbacks to this technology for protecting user geo-privacy. First, SIFT removal and injection attack may add distortion and artifacts to the image. In contrast to our approach, such technology cannot take for granted that the aesthetics of the image will be preserved.
	Second, such technology requires users to add a separate step to their photo-taking and image enhancement practices. 
	A certain number of users would likely adopt the technologies, but large-scale adoption would demand an unprecedented level of acceptance. 
	
	\subsubsection{Human-based solutions}
	
	So-called `human-based solutions' would require users to change their behavior in response to awareness of the risks of online photo sharing.
	Social networking and photo-sharing websites are increasingly offering the opportunities of sharing photos privately. 
	Researchers studying user behavior with respect to photo privacy have found that users experience a strong desire to participate in social sharing, which exerts a force on them to accept problems resulting from lack of control over their identity and disclosures~\cite{Besmer:2010}.
	The difficulty that users experience in dealing with complicated privacy settings is uncovered by \cite{Knijnenburg:2013}
	As a result of these difficulties, users make the choice of easy (i.e., public) sharing over privacy protective measures when they get overloaded.
	People are worried that the photo will not be shared properly and this impact their decision to share privately \cite{Ahern:2007}.
	Computer scientists have developed many mechanisms to protect users' privacy, such as encrypted email and messaging.
	However, users are slow to adopt such solutions for a range of reasons.
	%	Even when installation is made easily, it appears that when interacting socially, users are afraid of being stigmatized as `paranoid' if they adopt tools to protect themselves.
	We reject solutions that require users to radically change their behavior for the simple practical reason that any approach to geo-privacy that is not actively used by users is effectively worthless. 
	
	\subsubsection{Strength of simplicity}
	Recently, appreciation has arisen for simple, practical, filter-based approaches in service of privacy protection. 
	The movement, typified by~\cite{Ciftci_2015}, has been led by researchers investigating how private information in images can be protected from \emph{human} viewers.
	In contrast, in this work we investigate approaches to protect image content from \emph{algorithms}.
	However, like these researchers we recognize the value of simple approaches that are easy to apply in practice.
	Additionally, we point out the recent work that has been carried out on deceiving image classifiers that use deep neural networks~\cite{fooled}.
	In the context of privacy preservation, deceiving a classifier means protecting the content of an image.
	Here, instead of \emph{classification} we focus on \emph{retrieval} approaches that are used to infer image content. Such approaches may ultimately prove more difficult to defeat, since they require minimally one correct match, rather than a set of classifier training data.
	
	\section{Experimental Setup}
	In this section, we set the stage for our analysis of the interaction of automatic GLE technology with user behavior by describing the setup of our experiments.
	
	\subsection{Data set and evaluation}
	\label{sec:dataset}
	In order to ensure the validity and impact our experimental results, we need a data set that is representative in scale for the state of the art in social image geo-location, and that is also publicly available, making it possible for researchers who follow us to reproduce the experimental findings of this paper.
	We choose the MediaEval Placing Task 2013 Data Set~\cite{ME2013Over}\footnote{\url{http://www.st.ewi.tudelft.nl/~hauff/placingTask2013Data.html}}, a Creative Commons licensed set of images gathered from Flickr, which fulfills both of these criteria.
	Our experiments consist of two parts, GLE matching, and result analysis, and call for the application of both commercial tools and manual analysis.
	In order to ensure that we do not surpass the bonds of the resources at our disposal, we concentrate our experiments on a set of target images (also called, `query images') that were taken in a specific region, San Francisco, California.
	We choose our region to be highly dense, since it this way it can be considered representative of the state which many regions of the world can be expected to reach, as the number of images taken by users continues to grow in future years.
	Table~\ref{table:dataset}, reports the basic statistics on the composition of our data set.
	\begin{table}[h]
		\resizebox{\columnwidth}{!}{
			\begin{tabular}{|c|c|c|c|}
				\hline
				\textbf{\begin{tabular}[c]{@{}c@{}}Background \\ Collection\end{tabular}} & \textbf{Target Images} & \textbf{\begin{tabular}[c]{@{}c@{}}Toponym-Tagged\\ Target Images\end{tabular}} & \textbf{\begin{tabular}[c]{@{}c@{}}Toponym-Tagless\\ Target Images\end{tabular}} \\ \hline
				8,539,050                                                                 & 2,006                 & 1,264                                                                          & 742                                                                             \\ \hline
			\end{tabular}
		}	
		\caption{The total number of background collection images used in our experiments, together with statistics on our target images, which are from the San Francisco Area.}
		\label{table:dataset}
	\end{table}
	
	One aspect that we are invested in investigation in our experiments is the `photographic intent' of the user taking and sharing the image.
	As mentioned, in the introduction, we do not advocate `geo-cloaking' every social image, but rather giving the user control over geo-information that is shared along with social images.
	We are interested in making a separation between cases in which the user probably intended to reveal the location of the image, and cases in which the revelation of the location is more likely to be untended.
	This separation allows us to study the different impacts of image GLE and geo-cloaking for these two types of situation.
	We made the differentiation between images where the user probably did and possibly did not mean to reveal geo-information by analyzing the tags assigned by the uploader to the images.
	We put images that contain at least one place-related tag (i.e., toponym) into a set we refer to as \emph{toponym tagged images} and all other images are assigned to the set \emph{toponym tagless images}.
	To identify toponyms, we use an analysis of tag geo-distribution, described in~\cite{mediaeval2014}, to check how representative each tag is for location. Toponym detection is based on a concept to similar to the TF-IDF weighting scheme for words in a document collection. A tag is considered as a toponym when it appears frequently only around certain region, and is not considered a toponym if it appears in many different regions. As reported in Table~\ref{table:dataset}, 63\% of the target images had at least one toponym among their tags.
	%Note that the presence of toponym in the image's associated metadata is not necessarily sufficient to make a correct estimation. We used the presence of toponyms to divide the images into two sets of different intention. If at least one toponym is included in the image's tags, we assume that the uploader's intention include sharing of his/her whereabout, whereas if none of tags is  toponym (or if there is no tags at all), sharing the photo's location is not in user's intention.      
	
	The results of our experiments are reported in terms of two different kinds of evaluation.
	First, the evaluation performance of the image GLE system.
	This is measured in terms of the percentage of photos whose geo-locations were correctly estimated with a given distance threshold (i.e., radius) of their ground truth location.
	The ground truth of the photos is the position at which the photographer was standing when the picture was taken.
	We use the geo-coordinates recorded by the GPS of the user's camera, or assigned by hand to the user during the process of uploading to Flickr.
	We focus on two distance thresholds: 100 meters and 1 kilometer.
	We choose 100 meters since this small distance is most `dangerous' for malicious uses of image geo-locations, such as cybercasing, because it makes it possible to pick out the addresses of individual residences~\cite{Friedland:2010}.
	We choose 1km because this distance is also small enough to be quite worrisome for privacy considerations, but large enough to take into account the way in which uploaders add geo-coordinates to their photos.
	Some uploaders assign their photos the geo-location of the main object in the photo, rather that the photograph location. 
	The 1km distance allows us to take into account the fact that for some photos this distance is relatively far.
	
	Second, we report the performance the results of our experiments in terms of \emph{percent cloaked}.
	This is the proportion of the number of images that can be correctly geo-located with the original image, that can no longer be correctly geo-located once a specific `enhancement', e.g., a filter, has been applied.
	Note that in most GLE papers, the authors are attempted to increase the \emph{percent correct} of images whose location is correctly predicted. In this work, we are interested in user behaviors that \emph{decrease} the percent correct, and rather \emph{increase} the percent cloaked.

	\subsection{Visual geo-location estimation approach}
	%JC: need to finish this description of the baseline system
	Next we describe the GLE systems that were used for the experiments. Both systems use a top-one nearest neighbor approach to GLE. This means that the target photo is used as a query, and the system returns a list of results from the background collection that are ranked with respect to visual similarity to the query. The top-ranked image is taken as the `geo-propagator' and its coordinates are used to predict the location of the target image. The systems differ with respect to the way in which they compute visual similarities. Below, we describe each in turn.
	
	\subsubsection{Baseline: Basic Nearest Neighbor (BNN)}
	As a baseline for image GLE we choose a very basic system that implements the standard bag-of-visual-words paradigm for image retrieval in a large scale dataset~\cite{Jegou}.
	A codebook of 1000 visual words was generated using SIFT descriptors from randomly sampled subset of the background collection. 
	Randomized KD-trees were trained with the background collection images and
	%Query images are placed to the location of the top-1 nearest neighbor in the background collection. 
	vlfeat\footnote{\url{http://www.vlfeat.org/}} was used for extracting SIFT feature~\cite{LOWE}. 
	FLANN(Fast Library for Approximate Nearest Neighbor) \footnote{\url{http://www.cs.ubc.ca/research/flann/}} was used for nearest-neighbor matching. Images in the database that could not be handled by vlfeat was excluded and 7,563,041 images were used for generating index for the baseline system. 
	
	\subsubsection{Advanced: Pairwise Geometric Matching (PGM)}
	Our main system for image GLE represents the state-of-the-art in GLE for social images.
	For visual matching, it exploits a pairwise geometric matching technique, recently introduced in~\cite{PGMCVPR15}. 
	Like the BNN system, it built upon the standard bag-of-visual-words paradigm, which scales up well to a large-scale datasets~\cite{ThreeThingsCVPR12}. 
	
	To speed up retrieval and also improve accuracy, we also adopt the technique which was proposed~\cite{HEJ2010}, which represents subregions of the feature space by signatures, and compares descriptors based, not only on visual words, but also on the distance between their subregions within the feature space. When calculating the initial ranking score, we employ the burst weighting scheme developed in~\cite{burstinessCVPR09}. 
	In order to deal with quantization noise introduced by visual word assignment, we take advantage of the strategy used in~\cite{HEJ2010, burstinessCVPR09}.
	The strategy assigns a given descriptor to several nearest visual words. 
	Since multiple assignment strategy causes a significant increase in the number of visual words per image, e.g., on average 4.2 visual words per descriptor, we apply it only to the query image, and not to the images in the background collection. 
	
	To find the reliable correspondences, the initial list is re-ranked using a geometric matching techniques.
	Our technique, pairwise geometric matching, improves over existing techniques because it not only requirements matches between key points, it also requires matches between pairs of matching key points. The details of the algorithm, and a series of experiments which demonstrate its superiority to existing approaches, are described in~\cite{PGMCVPR15}. We deploy the BoofCV software to extract SURF descriptors with default parameters and use exact k-means to cluster these descriptors and generate visual words. To mimic the situation in a real retrieval system, we use a separate set of $50k$ randomly selected images from Flickr to train the \textit{generic} $20k$ vocabulary set and use it in all experiments. 
	
	\section{Impact of photo-taking behavior}
	We now turn to our experiments and analysis. In this section, we investigate how the behavior of the photographer impacts the ability of image GLE to predict correct geo-locations for photos. We are specifically interested in how the photographers choices of framing and location impact whether the photo can be correctly geo-located. We also examine
	
	\subsection{Geo-locatable social images}
	First, we discuss the overall performance of our image GLE systems.
	Table~\ref{table:overall_result} shows the performance of two visual geo-location estimation system. BNN-SIFT represents the baseline nearest neighbor approach using SIFT descriptors and PGM-SURF represents the pairwise geometric matching approach using SURF descriptors. Each number represents the percentage of query images those geo-locations are estimated within the distance thresholds from the groundtruth. 
	%Since the PGM-SURF employs many state-of-the-art approaches in the image retrieval domain, it outperforms BNN-SIFT in all settings of the experiments as expected. This reflects the fact that the state-of-the-art has moved forward very quickly. 
	We see that, as expected, the advanced PGM-SURF system outperforms the baseline BNN-SIFT system, and is able to correctly estimate the location of nearly 12\% of the photos in the test collection.
	
	% Overall performance table here
	\begin{table}[!h]
		\centering
		\resizebox{\columnwidth}{!}{
			\begin{tabular}{|c|c|c|c|c|c|c|}
				\hline
				\textbf{}         & \multicolumn{2}{c|}{\textbf{Overall}} & \multicolumn{2}{c|}{\textbf{Toponym-Tagged}} & \multicolumn{2}{c|}{\textbf{Toponym-Tagless}} \\ \hline
				\textbf{}         & \textbf{100m}      & \textbf{1km}     & \textbf{100m}         & \textbf{1km}         & \textbf{100m}          & \textbf{1km}         \\ \hline
				\textbf{BNN-SIFT} & 3.04               & 8.13             & 3.32                  & 9.33                 & 2.56                   & 6.06                 \\ \hline
				\textbf{PGM-SURF} & 7.63               & 11.86            & 9.49                  & 12.10                & 4.45                   & 11.46                \\ \hline
			\end{tabular}
		}
		
		\caption{Geo-location estimation performance of two systems: Each number represents the percentage of query photos that are correctly estimated within the distance thresholds 100m and 1km from the groundtruth.}
		\label{table:overall_result}
	\end{table}

	Note that although this is the first time that PGM results have been reported for a problem at this scale, the focus here is not on the image GLE algorithm itself, but rather on widespread user photo-taking and enhancement practices. For this reason, it is important that the PGM algorithm is state of the art, but we are not looking to improve it, but rather we are investigating the factors that negatively impact its performance.	
	When compared to the ultimate target of image GLE, to correctly geo-locate 100\% of social images, our advanced system, which can only geo-locate 12\% of images seems rather unthreatening. 
	Unfortunately, this feeling is a false sense of security.
	Recall from above, that geo-privacy is compromised by large scale mining of images that searches for vulnerable victims to target.
	For people who leak visual geo-information by mistake, the potential of 12\% to translates into direct damage is something that the community needs to take seriously, given that in 7.63\% of the cases, the information is exact to the 100 meters level.
	Further, the comparison with the baseline shows that the technology is rapidly improving, suggesting that the coming years will see further improvements.
	The current state of GLE technology is a strong motivator to start, as is our goal in this paper, on the large mission of understanding how control over visual geo-information can be put in the hands of this paper.
	\vspace{3mm}
	\subsection{Photographer geo-intent}
	As mentioned in the introduction, our focus is on examining the geo-privacy aspects of images for which the user may not be aware that GLE can automatically predict their location.
	As described in Section~\ref{sec:dataset}, we divide the target image set into ``toponym-tagged'' and ``toponym-tagless'' subsets since we are interested is studying cases separately in which the photographer probably was and may not have been aware that their photos were communicating geo-information. 
	The results in Table~\ref{table:overall_result} demonstrate that it is easier for estimate the geo-location of "toponym-tagged" images rather than "toponym-tagless" images. This result suggests that when the photographer has a true `geo-intent', i.e., intends to reveal the location of the photo, that location is also better reflected in the visual content of the photo, than it is in cases in which the photographer does not necessary want to communicate location along with the photo.
	
	There is apparently a connection between not necessarily wanting the location of a photo to be known, and not including visually geo-specific content in the photo that would make the photo findable.
	This result provides a first indication that we are on the right track with our idea that certain user behaviors `naturally' protect images.
	Understanding these users behavior, will, ultimately make it possible to reinforce them, reducing the incidences of geo-information leaks compromising user privacy.
	
	\subsection {Location and collective photo patterns}
\begin{figure}[h]	
		\includegraphics[clip, width=\columnwidth]{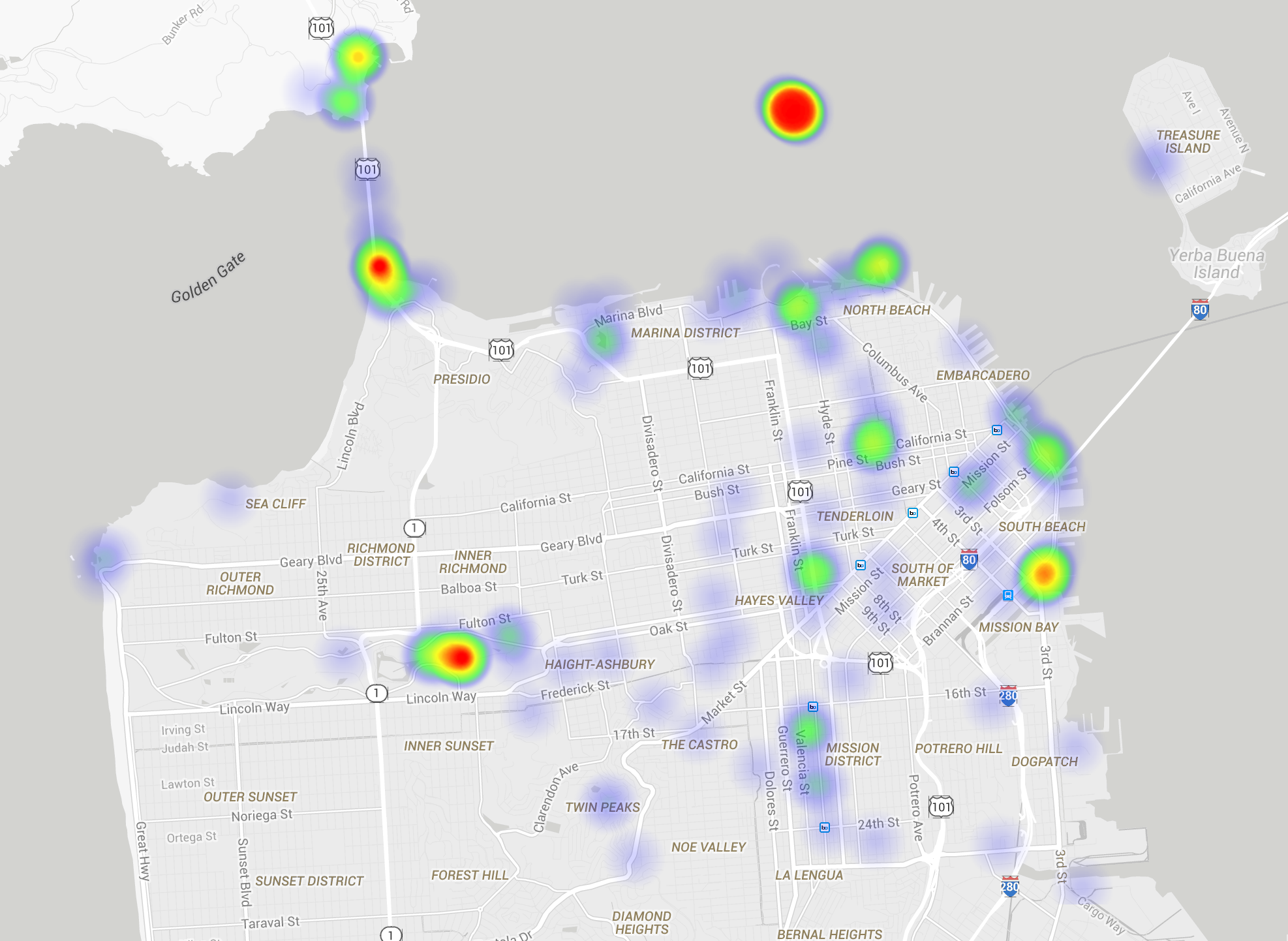}
		
		\caption{Where to be wary: Heat map showing the distribution of the photos in our test set of target images whose geo-locations can automatically be correctly predicted.}
		\label{fig:heat}
	\end{figure}

	Next, we turn to consideration of how the collective photo-taking behavior of users impacts automatic image GLE. Recall that the `GLE conditions' involve an image matching other images taken at its location more closely than it matches images taken at different locations. For this reason, we go beyond considering whether individual images can be geo-located, to also studying the geo-location properties of the collection as a whole. Our first step is to visualize our area of interest in terms of locations at which users' geo-privacy is particularly vulnerable to compromise (Fig.~\ref{fig:heat}).

	These patterns result from the fact that certain places are more frequented by people, because they are interesting in and of themselves, or that events that are interesting take place here.
	We are unsurprised to note that photographs at touristic destinations such as the Golden Gate bridge and Alcatraz. However, the map makes in Fig.~\ref{fig:heat} makes clear that image GLE technology goes beyond touristic landmarks in its ability to predict the geo-locations of photos. As such, it is not reasonable to assume that a users can develop intuitions for where they are in a `safe' place to take a photo without compromising their geo-privacy. Looking towards the future, we note that these patterns of geo-distribution can be expected to change as collections like Google street view become more readily available, for example, the data set used by ICMLA Streetview Recognition Challenge~\cite{icmla_challenge}. 
	
	\subsection {Photographer framing}	
	Now, we turn to investigate the question of whether there are certain, common behaviors of photographers that contribute to images being naturally protected from automatic GLE. Here, we carried out a qualitative investigation of a selection of the `hot spots' in San Francisco, shown in Fig.~\ref{fig:heat}, comparing the images that were correctly geo-located with those that were not. We are interested in photographic framing, which includes the way in which the photographer composes the image, but also includes other decisions about how the image is taken, including lighting~\cite{Riegler:2014}. Our investigation, which was carried out by manual analysis, revealed a set of interesting factors with a geo-cloaking effect on images. Specifically, we discovered three photographer behaviors that made the difference between a photo being geo-located or geo-cloaked: use of unexpected angles, use of unconventional scales, taking photos at different times of the day.
	
	\begin{figure}[t]	
		\includegraphics[clip, width=\columnwidth]{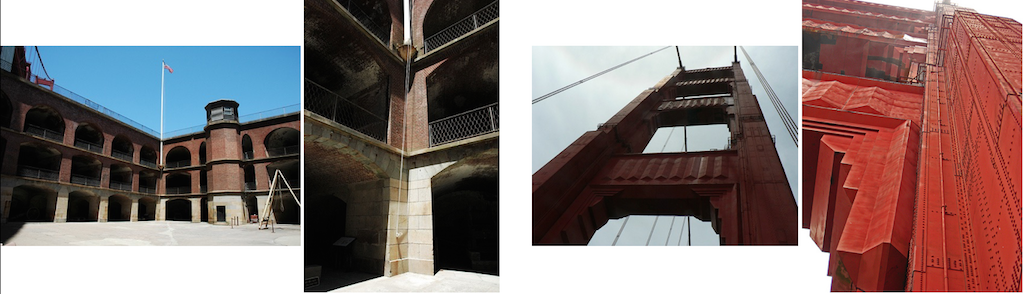}
		
		\caption{Zoomed (or scaled) image may have wrong GLE despite the existence of images with a wider view in the background collection.}
		\label{fig:scale}
	\end{figure}	
	
	Fig.~\ref{fig:firstimage} illustrates cases in which the angle (top row) and the time of day (middle row) differentiate a photo that can be automatically geo-located. In Fig.~\ref{fig:scale}, we show the case of zoomed in photos being protected. Note that these effects cannot be attributed to the underlying visual features, which are rotation, lighting, and scale invariant. 
	Rather, they are due to the fact that diverse framing means that photos have less visual overlap with each other, and for this reason are less likely to have a `geo-propagator' in the background collection that can be used to correctly predict their geo-location.
	The evidence points to the conclusion that the drive of photographers to use different types of framing to make their photos interesting and attractive naturally protects their photos from automatic GLE.
	As more and more photos are put online, a specific framing of a scene may be picked up and used by multiple photographers.
	However, if the choice of framing is made by photographers because they are attempting to create unique images, then we can anticipate a trend towards natural diversification, and the continued contribution of framing effects to balancing geo-location with geo-cloaking.
	\newline
	\section{Impact of image enhancement}
	In this section, we move beyond the act of photo-taking to also examine practices of image enhancement that are in widespread use among users who share photos online. We  investigate how enhancement affects automatic GLE of individual photos, and also how enhancement practices scale up as they are used not just by individual users, but by many users contributing images to the background collection.
	
	\subsection{Users behavior and image enhancement}
	We define image enhancement as filters and other edits that people apply during or after image capture in order to heighten the attractiveness of the image.
	Specifically, we are interested in enhancements that are highly popular, since these best reflect the widespread natural user behavior, which is the focus of our study.
	We focus on filters and features that are available to users on Instagram\footnote{\url{https://instagram.com/}}, one of the most widely used photo sharing/social networking services.
	
	Perhaps the simplest thing that users do to make their images more attractive is cropping. Cropping is often used to cut out clutters in the image and helps the viewer to concentrate  on the main subject. Images evidently lose some of its contents along the edge lines when cropped. We are interested in cropping because of its potential to cut out incidental matches in background that could allow photos to be geo-located. Next, we are interested in tilt-shift. Tilt-shift photography manipulates focus and depth-of-field to yield photos that have selective focus, often mimicking a miniature model. Instagram provides a post-production fake tilt-shift feature. Unlike cropping, tilt-shifted image still retains all of the original pixels, although the areas outside the focus point becomes blurred. We are interested in tilt-shift because it is popular, and also because the blur potentially eliminates matches in background.
	
	We focus a large number of experiments on filters. A filter is a preset recipe for adjusting various photographic settings (exposure, contrast, color balance, etc) of a photo to give it certain look and feel. 
	Instagram offers a large number of filters that are carefully tailored, and provide users with a convenient means of enhancing the attractiveness of the photo, with no need for special skill. 
	We are interested in Instagram filters, since their popularity strongly suggests that users feel that they add value to images.
	Effectively, these filters represent a common practice already applied by users that potentially results in a loss of visual information in images, that can provide a geo-cloaking effect.
	
	%For instance, `Kelvin' filter increases saturation and temperature to give the photo a radiant `glow'. 
	
	%	There are many more visual editing features provided by Instagram. 
	%	Among them, we tried to pick the features that is visually pleasing but lossy that image retrieval performance will degrade. 

	%How the experiments were carried out (including the specific enhancements that were chosen, how they were applied).
	
\begin{figure}[!t]	
		\includegraphics[clip, width=\columnwidth]{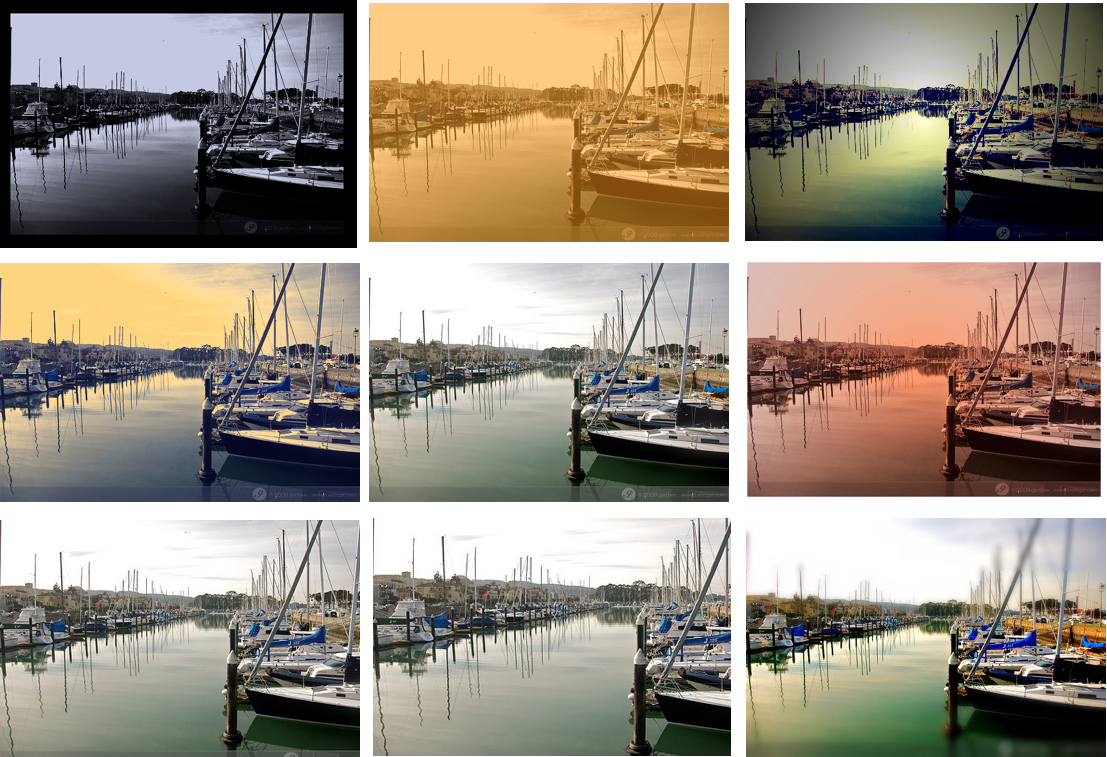}
		
		\caption{Image enhancements (popular on Instagram) use in our experiment. Center is the original image. Clockwise from top left: Gotham, Kelvin, Lomo, Toaster, Tiltshift, Crop40\%, Crop20\% and Nashville.}
		\label{fig:effect_example}
		
	\end{figure}

	Instagram currently provides 27 filters for use. According to the survey released by TrackMaven~\cite{trackmaven}, normal or \#nofilter is by far the most popular filter with 26\% of the total photos.  Among the popular filters are Kelvin, Lomo, Nashville, and Toaster, which we selected for our study because they have a relatively strong influence on the original image, and thereby promising geo-cloaking effects. Although the Gotham filter has been discontinued by Instagram, we chose to also include it in our study, since effects are very dramatic with lots of contrast added. The following gives an overview of the filters. Figure~\ref{fig:effect_example} shows an example with each filters' effect. 
	
	\begin{itemize}[noitemsep]
		\item Gotham filter produces a black\&white, high contrast image with bluish undertones.
		\item Kelvin filter applies a strong peach/orange overlay. 
		\item Lomo filter boosts contrast, and applies vignette (i.e., reduction in brightness/saturation around the edges). 
		\item Nashville filter gives a washed out 80s fashion photo feel with warmed up temperature, lowered contrast and increased exposure. 
		\item Toaster filter features vivid colors with pink/orange glow out of the center. 
	\end{itemize}
	
	To carry out the experiments we applied the enhancements to the target photos, and, for the final experiment, also to images in the background collection. 
	Instagram filters were applied using ImageMagick\footnote{\url{http://www.imagemagick.org}} in a Python wrapper.
	Cropping was done using Smart Cropping service provided by Imagga\footnote{\url{https://imagga.com/technology/smart-cropping-and-slicing.html}}. 
	The Imagga cropping API analyzes images in terms of composition, color and object localization, and attempts to crop in a way that the main subject material of interest remains in the photo.
	Cropping was applied to crop out 20\% and 40\% of the original image, respectively. The aspect ratio of the photo was preserved.
	Tilt-shift effect was applied with a simple Python script~\cite{python-tilt-shift}. We took the center point of a cropped image generated by Smart Cropping as the focus point of the original image, and the y-axis value of the center point was used to apply linear tilt-shift. 
	
	\subsection{Geo-cloaking effects of image enhancement}
	Our experiments revealed that our enhancement filters demonstrated a clear ability to protect the location of images from being automatically predicted by GLE. An overview of the geo-cloaking performance of the filters is provided in Table~\ref{tab:geo-compare}.

	\begin{table}[h]
		\resizebox{\columnwidth}{!}{
			\begin{tabular}{|c|c|c|c|c|c|c|c|c|}
				\hline
				\multirow{2}{*}{} & \multicolumn{5}{c|}{\textbf{Filter}}                                     & \multicolumn{2}{c|}{\textbf{Crop}} & \multirow{2}{*}{\textbf{\begin{tabular}[c]{@{}c@{}}Tilt\\ shift\end{tabular}}} \\ \cline{2-8}
				& \textbf{GOT} & \textbf{KEL} & \textbf{LOM} & \textbf{NAS} & \textbf{TOA} & \textbf{20\%}    & \textbf{40\%}   &                                                                                \\ \hline
				\textbf{SIFT-BNN} & 64.1         & 12.8         & 44.9         & 15.4         & 75.8         & -1.2             & 5.8             & 21.5                                                                           \\ \hline
				\textbf{SURF-PGM} & 51.0         & -2.6         & 29.4         & 5.9          & 53.6         & -3.3             & 4.6             & 15.7                                                                           \\ \hline
			\end{tabular}
		}
		\caption{Net geo-cloaking percentage: Ratio between the number of images that can be automatically geo-located after and before an enhancement has been applied. (Filters: GOTham, KELvin, LOMo, NAShville, TOAster. Crop 20\% and 40\% represents the percent of the image cropped.)}
		\label{tab:geo-compare}
	\end{table}
	
\begin{figure}[!h]
		\begin{center}
			\includegraphics[clip, width=\columnwidth]{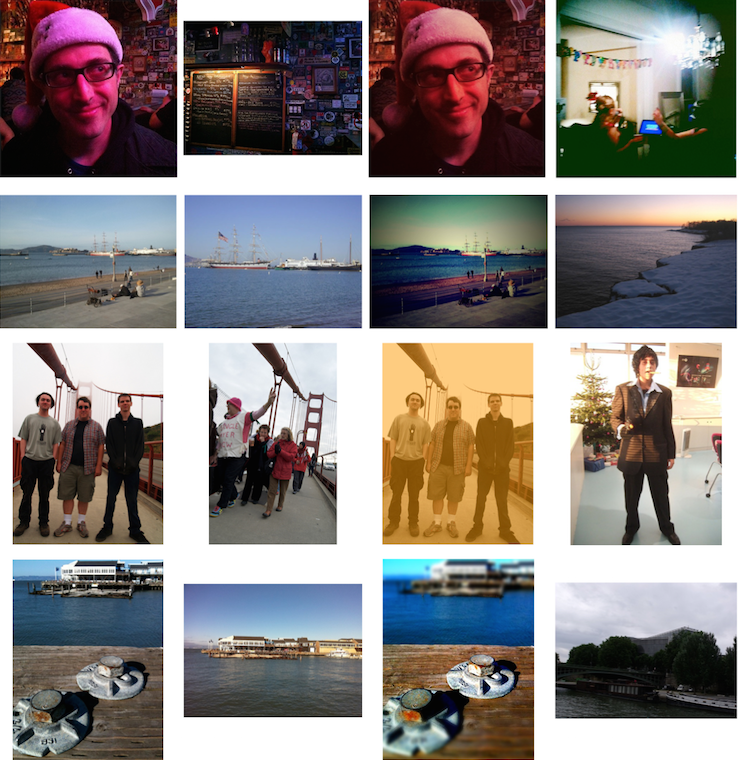}
		\end{center}
		\caption{Example images whose locations are protected \emph{after} enhancement. First column is the original image; second column is the `geo-propagator' that allows the original image to be geo-located; third column is the enhanced image; fourth column is the nearest neighbor of the enhanced image under PGM, which no longer propagates the correct coordinates. From top to bottom, the enhancements are: Toaster, Lomo, Kelvin, tilt-shift.}
		\label{fig:cloak_effect}
	\end{figure}

	These results are surprising, since they clearly show that users' natural behavior can serve to help protect them from geo-privacy leaks, without requiring them to adopt new, yet-unknown technology or applications, or radical changes in their current photo sharing practices. It is easier to protect images against the basic nearest neighbor approach, with the Nashville filter protecting 75\% of all images that were originally correctly geo-located by BNN. However, even against the more powerful PGM approach, many filters do impressively well. These findings suggest that geo-distinctive information is being naturally removed from the photo, as the user filters the photo to enhance its attractiveness. Examples of the protective effect of enhancements at work are provided in Fig.~\ref{fig:cloak_effect}.
	
\begin{figure}
		\begin{center}
			\includegraphics [scale=0.35]{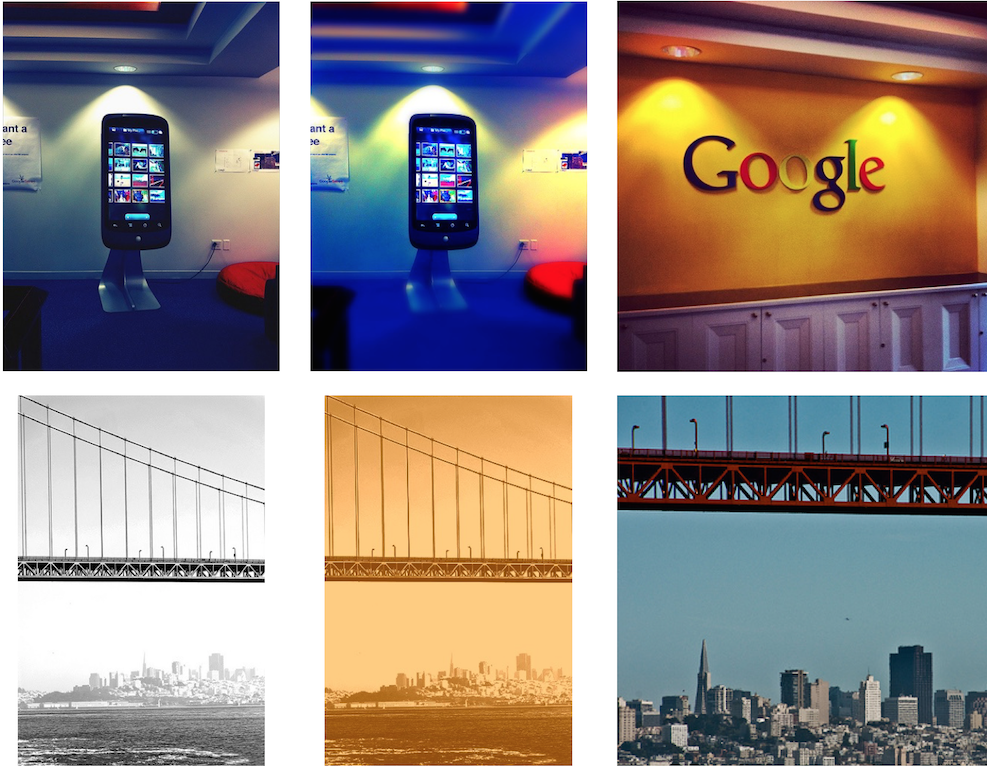}
		\end{center}
		\caption{Example images whose locations were protected \emph{before} enhancement. The first column is the original image, which could not be geo-located; the second is the enhanced image; and the third column is the geo-propagator matched to the enhanced image (but not the original). Enhancements are: Tilt-shift (top row) and Kelvin (bottom row).}
		\label{fig:reveal_effect}
	\end{figure}

	The interesting exceptions are the cases in which enhancement serves to open images to risk. The negative values in Table~\ref{tab:geo-compare} are the cases in which an enhancement actually reveals the locations of more images than it cloaks. Kelvin apparently promotes automatic GLE by exposing contours of objects in images, and cropping helps to set the focus on the main subject of the photo, which is apparently often geo-distinctive. Examples in which enhancements allow the location of a photo to be automatically predicted when the location of the original was not are shown in Fig.~\ref{fig:reveal_effect}. In the top row, the original (left) is blurred by tilt-shift (middle) making is similar to a correct `geo-propagator', and less similar to other images in the background collection containing horizontal lines at the top. In the second row, the contrast of the original (left) is enhanced by Kelvin (middle) making it more similar to images of the correct bridge, and less to other similar hanging bridges in the collection.
	
	In order to understand these results in more detail, we display an additional table devoted exclusively to PGM. Here, we report the results in terms of the percentages of photos whose geo-location has been correctly predicted. Results are shown at both 100m and 1km, and also broken down into the ``Toponym-tagged" and ``Toponym-tagless" cases.

	\begin{table}[!h]
		\centering
		\resizebox{\columnwidth}{!}{
			
			\begin{tabular}{|c|l|c|c|c|c|c|c|c|c|c|}
				\hline
				\multicolumn{2}{|c|}{} &  & \multicolumn{5}{c|}{{\bf Filter}} & \multicolumn{2}{c|}{{\bf Crop}} &  \\ \cline{4-10}
				\multicolumn{2}{|c|}{\multirow{-2}{*}{{\bf }}} & \multirow{-2}{*}{{\bf \begin{tabular}[c]{@{}c@{}}No \\ Filter\end{tabular}}} & GOT & KEL & LOM & NAS & TOA & 20\% & 40\% & \multirow{-2}{*}{{\bf \begin{tabular}[c]{@{}c@{}}Tilt\\ Shift\end{tabular}}} \\ \hline
				\multicolumn{2}{|c|}{} & 7.63 & 3.74 & 7.83 & 5.38 & 7.18 & 3.53 & 7.28 & 7.88 & 6.43 \\ \cline{3-11} 
				\multicolumn{2}{|c|}{\multirow{-2}{*}{{\bf Overall}}} & \cellcolor[HTML]{C0C0C0}11.86 & \cellcolor[HTML]{C0C0C0}6.48 & \cellcolor[HTML]{C0C0C0}11.86 & \cellcolor[HTML]{C0C0C0}8.23 & \cellcolor[HTML]{C0C0C0}10.97 & \cellcolor[HTML]{C0C0C0}5.63 & \cellcolor[HTML]{C0C0C0}11.76 & \cellcolor[HTML]{C0C0C0}12.11 & \cellcolor[HTML]{C0C0C0}10.22 \\ \hline
				\multicolumn{2}{|c|}{} & 9.49 & 4.83 & 9.34 & 5.78 & 7.99 & 3.88 & 8.62 & 8.94 & 7.12 \\ \cline{3-11} 
				\multicolumn{2}{|c|}{\multirow{-2}{*}{{\bf \begin{tabular}[c]{@{}c@{}}Toponym-\\ tagged\end{tabular}}}} & \cellcolor[HTML]{C0C0C0}12.1 & \cellcolor[HTML]{C0C0C0}7.83 & \cellcolor[HTML]{C0C0C0}12.82 & \cellcolor[HTML]{C0C0C0}10.28 & \cellcolor[HTML]{C0C0C0}11.95 & \cellcolor[HTML]{C0C0C0}7.36 & \cellcolor[HTML]{C0C0C0}13.45 & \cellcolor[HTML]{C0C0C0}14.16 & \cellcolor[HTML]{C0C0C0}11.08 \\ \hline
				\multicolumn{2}{|c|}{} & 4.45 & 1.89 & 5.26 & 4.72 & 5.80 & 2.96 & 4.99 & 6.06 & 5.26 \\ \cline{3-11} 
				\multicolumn{2}{|c|}{\multirow{-2}{*}{{\bf \begin{tabular}[c]{@{}c@{}}Toponym-\\ tagless\end{tabular}}}} & \cellcolor[HTML]{C0C0C0}11.46 & \cellcolor[HTML]{C0C0C0}4.18 & \cellcolor[HTML]{C0C0C0}10.24 & \cellcolor[HTML]{C0C0C0}4.72 & \cellcolor[HTML]{C0C0C0}9.30 & \cellcolor[HTML]{C0C0C0}2.70 & \cellcolor[HTML]{C0C0C0}8.89 & \cellcolor[HTML]{C0C0C0}8.63 & \cellcolor[HTML]{C0C0C0}8.76 \\ \hline
			\end{tabular}
		}
		\caption{Geo-location estimation results: Percentage of target images correctly geo-located by PGM-SURF for the enhancements GOTham, KELvin, LOMo, NAShville, TOAster, cropping and tilt-shift. White and gray-scaled cells show results at 100m and 1km, respectively.}
		\label{table:SURF_result}
	\end{table}
	
	Two major insights can be gained from this table. First, the ability of enhancements to geo-cloak photos is comparable at both the 100m as well as the 1km level, although cloaking is slightly more effective at 100m. Second, enhancements help to geo-cloak both `toponym-tagged' and `toponym-tagless'. However, interestingly only certain filters are helpful for `toponym tagless' photos. Lomo, Nashville and 20\% cropping actually make it easier for GLE to predict the location of ``toponym tagless'' photos. In particular, we note that the bottom two rows of  Fig.~\ref{fig:firstimage} contain cases in which the original photo not only fails to have a ``toponym tag'', if fails to have any tag at all. At the same time, the visual match between these photos involves very small areas in their background. Apparently, filters act differently in making these areas more salient. In any case, we interpret this result as signal that there is an underlying difference between photos that are more or less likely to be intentionally linked to their geo-locations by the photographer, which is worthy of future study.

	\vspace{3mm}
	\subsection{Collective image enhancement patterns}
	Our experiments so far have yielded the satisfying insight that has determined that the use of image enhancement filters can lead to the substantial reduction of the ability of a geo-estimation system to correctly predict the location of an image.
	However, an obvious question to ask is whether enhancements might lose their power to protect geo-privacy as the number of enhanced images in the background collection grows.
	Our final experiment takes a look at what happens as background images are themselves enhanced.
	In real life, there are two considerations that would impact how filtered images would accumulate in the background collection: first, the number of images that is filtered overall, and second, the number of images that are filtered with a given type of filter.
	Here, we chose to focus on a scenario that constitutes a `worse case' for geo-cloaking. 
	In particular, we investigate the effects when the same enhancement is applied to both the target image and also specifically to potential `geo-propagators' in the background collection.
	We perform this experiment using the BNN-SIFT system. The top-100 visual nearest neighbors of the original target photo are retrieved.
	Then both the target photo and the top-100 list are cloaked with a the same filter.
	The enhanced top-100 images are then introduced into the background as replacement for the original background collection images.
	We then carry out GLE as usual.
	
	Tab.~\ref{tab:collection} reports the percentages of enhanced target photos that were retrieved when their nearest neighbors were also enhanced.

	\begin{table}[h]
		
		\resizebox{\columnwidth}{!}{
			\begin{tabular}{|c|c|c|c|c|c|c|}
				\hline
				\multicolumn{2}{|c|}{}                                                                                               & \textbf{GOT} & \textbf{KEL} & \textbf{LOM} & \textbf{NAS} & \textbf{TOA} \\ \hline
				\multirow{2}{*}{\textbf{\begin{tabular}[c]{@{}c@{}}Original\\ Back. Collect. \end{tabular}}} & \textbf{100m} & 1.51         & 3.08         & 2.15         & 2.88         & 1.41         \\ \cline{2-7} 
				& \textbf{1km}  & 4.48         & 8.14         & 5.3          & 7.21         & 3.91         \\ \hline
				\multirow{2}{*}{\textbf{\begin{tabular}[c]{@{}c@{}}Filtered\\ Back. Collect. \end{tabular}}} & \textbf{100m} & 1.33         & 2.8          & 1.96         & 2.53         & 1.22         \\ \cline{2-7} 
				& \textbf{1km}  & 4.08         & 7.56         & 4.76         & 6.21         & 3.56         \\ \hline
			\end{tabular}
		}
		\caption{Comparison of percentage of enhanced target images whose location is correctly predicted using the original collection, and using a collection in which the same enhancement has been applied to the visual nearest neighbors.}
		\label{tab:collection}
	\end{table}
	
	Overall, the results of this experiment point to the conclusion that a growth in the use of filters will not reduce their ability to geo-cloak users' photos.
	In fact, enhancing the collection slightly reduces the ability of the GLE approach to correctly predict the location of images.
	With these results we are apparently observing that filters introduce a loss of that sort of information that is important in order for automatic GLE approaches to function optimally.
	
	\section{Visual geo-privacy challenges}
	We have argued in this paper that the current state of geo-location estimation technology warrants concern about geo-privacy of users who take photos and share them online. 
	The conditions under which the location of a photo can be automatically estimated depend both on the photo, and on the contents of the data available to use as the background collection.
	Since these factors extend beyond what a user can be expected to be aware of, it is impossible for individual users to assess the risks to their own geo-privacy.
	Instead, they need support from technologies developed by visual information retrieval experts.
	
	Our ultimate goal is to give users more control over geo-privacy by supporting them in preventing unintended leaks of geo-information while sharing photos online. Simple solutions, such as `No-sharing', fail because of the immediacy of other concerns during the photosharing process. Even if users share only privately, photos can fall into the wrong hands.
	One part of the solution is obviously educating users, and a growing number of efforts are under way to actually teach users about social multimedia privacy~\cite{teachpriv}. However, an important part of the solution involves technical challenges.

	This paper has pointed out that although geo-privacy concerns are growing, the situation is not hopeless.
	Rather there are valuable insights that can be provided by research in the area of visual information retrieval.
	We have shown that users' own photo taking and image enhancement behavior actually has a contribution to make to protecting their privacy.
	We have systematically evaluated popular practices in photo-taking and image enhancement, and shown their impact on state-of-the-art geo-location estimation technology.
	Although certain framing practices and Instagram filters are effective in helping to protect privacy, there is no silver bullet solution. Instead, more work is needed to understand the interactions of photo-taking behavior and enhancements.
	
	In closing, we offer a concrete proposing for a research goal. Moving forward GLE research should strive towards the vision of the \emph{geo-privacy aware camera}. Such a camera would use visual information retrieval techniques to alert users in real time when they take a picture that potentially leaks their geo-location via its visual content. This device is non-trivial, since it needs to access a enormous, and also growing number of background images to estimate the vulnerability of a new image. However, the estimation times achieved by our PGM approach are reasonably fast to be useful to users. Using a Map-Reduce structure on a Hadoop distributed server, our system achieves an average response time of 1.8 seconds per target image. Research dedicated, for example, to techniques that approximate the background collection, could take a step nearer to fitting the collection in memory, and thereby a step nearer to running a geo-privacy aware camera from a mobile phone. We envision such an application would also recommend to the user filters that add Instagram style enhancements, but that are specially chosen to protect geo-privacy. The key challenge of the geo-privacy aware camera is that it should fit seamlessly into currently widespread photographic practices, i.e., phototaking behaviors and the use of image enhancements.
	
	Looking towards the future, it is critical that research focus not only on improving performance of automatic GLE, but also on maintaining the \emph{correct balance} between research on technologies that predict the geo-location of social images, and technologies that protect geo-privacy.

	%ACKNOWLEDGMENTS are optional
	%\section{Acknowledgments}
	%<anonymized>

	%acknowledgment 
	%	This research is supported in part by NSF grant 1065240 ``TC:Medium:Understanding and Managing the Impact of Global Inference on Online Privacy'' and NSF grant 1251276, ``BIGDATA: Small: DCM: DA: Collaborative Research: SMASH: Scalable Multimedia content AnalysiS in a High-level language''.
	
	%
	% The following two commands are all you need in the
	% initial runs of your .tex file to
	% produce the bibliography for the citations in your paper.
	\bibliographystyle{abbrv}
	\bibliography{sigproc} %, refsall}  % sigproc.bib is the name of the Bibliography in this case
	% You must have a proper ".bib" file
	%  and remember to run:
	% latex bibtex latex latex
	% to resolve all references
	%
	% ACM needs 'a single self-contained file'!
	%
	%APPENDICES are optional
	%\balancecolumns
\end{document}